\documentclass[11pt]{article}
\pdfoutput=1  
\usepackage{graphicx,color}
\usepackage{appendix}
\usepackage{latexsym,amsmath,amssymb,graphicx,booktabs}
\usepackage{epsfig,latexsym,cite}
\usepackage{hyperref}
\numberwithin{equation}{section}

\definecolor{MyBlue}{rgb}{0.15,0.15,0.70}

\hypersetup{
colorlinks=true,
citecolor=MyBlue,
linkcolor=MyBlue,
urlcolor=MyBlue
}

\input epsf
\usepackage{amssymb}
\usepackage{amsmath}
\usepackage{amsfonts}
\usepackage{upgreek}
\usepackage{latexsym}

\newcommand{\nn}{\nonumber}

\newcommand{\iBox}{\Box^{-1}}

\renewcommand\({\left(}
\renewcommand\){\right)}
\renewcommand\[{\left[}
\renewcommand\]{\right]}

\newcommand\n{{\mbox {\boldmath $\nabla$}}}
\newcommand{\ra}{\rightarrow}

\def\lsim{\raise 0.4ex\hbox{$<$}\kern -0.8em\lower 0.62
ex\hbox{$\sim$}}

\def\gsim{\raise 0.4ex\hbox{$>$}\kern -0.7em\lower 0.62
ex\hbox{$\sim$}}

\def\lbar{{\hbox{$\lambda$}\kern -0.7em\raise 0.6ex
\hbox{$-$}}}

\newcommand\eq[1]{eq.~(\ref{#1})}
\newcommand\eqs[2]{eqs.~(\ref{#1}) and (\ref{#2})}
\newcommand\Eq[1]{Equation~(\ref{#1})}

\newcommand\eqss[3]{eqs.~(\ref{#1}), (\ref{#2}) and (\ref{#3})}
\newcommand\eqsss[4]{eqs.~(\ref{#1}), (\ref{#2}), (\ref{#3})
and (\ref{#4})}

\newcommand\eqst[2]{eqs.~(\ref{#1})--(\ref{#2})}
\newcommand\Eqst[2]{Eqs.~(\ref{#1})--(\ref{#2})}
\newcommand\pa{\partial}
\newcommand\p{\partial}

\newcommand\ee{\end{equation}}
\newcommand\be{\begin{equation}}
\def\bea{\begin{array}}
\def\eea{\end{array}}\def\ea{\end{array}}
\newcommand\ees{\end{eqnarray}}
\newcommand\bees{\begin{eqnarray}}
\def\nn{\nonumber}

\def\a{\alpha}
\def\b{\beta}

\def\g{\gamma}

\def\d{\delta}

\def\dslash{\hspace{-1mm}\not{\hbox{\kern-2pt $\partial$}}}
\def\Dslash{\not{\hbox{\kern-4pt $D$}}}
\def\pslash{\not{\hbox{\kern-2.1pt $p$}}}
\def\kslash{\not{\hbox{\kern-2.3pt $k$}}}
\def\qslash{\not{\hbox{\kern-2.3pt $q$}}}

\newcommand{\vac}{|0\rangle}
\newcommand{\cav}{\langle 0|}

\newcommand{\vk}{{\bf k}}
\newcommand{\vx}{{\bf x}}

\def\p1{{\bf p}_1}
\def\p2{{\bf p}_2}
\def\k1{{\bf k}_1}
\def\k2{{\bf k}_2}

\newcommand{\emn}{\eta_{\mu\nu}}

\newcommand{\gmn}{g_{\mu\nu}}

\newcommand{\gMN}{g^{\mu\nu}}

\newcommand{\gbmn}{\bar{g}_{\mu\nu}}

\newcommand{\pam}{\pa_{\mu}}

\newcommand{\pan}{\pa_{\nu}}
\newcommand{\parho}{\pa_{\rho}}

\newcommand{\paM}{\pa^{\mu}}
\newcommand{\paN}{\pa^{\nu}}
\newcommand{\paR}{\pa^{\rho}}
\newcommand{\paS}{\pa^{\sigma}}

\newcommand{\Rmn}{R_{\mu\nu}}
\newcommand{\Gmn}{G_{\mu\nu}}
\newcommand{\RMN}{R^{\mu\nu}}

\newcommand{\Rmnrs}{R_{\mu\nu\rho\sigma}}

\newcommand{\Tmn}{T_{\mu\nu}}
\newcommand{\Smn}{S_{\mu\nu}}

\newcommand{\dddM}{\kern 0.2em \raise 1.9ex\hbox{$...$}\kern -1.0em \hbox{$M$}}
\newcommand{\dddQ}{\kern 0.2em \raise 1.9ex\hbox{$...$}\kern -1.0em \hbox{$Q$}}
\newcommand{\dddI}{\kern 0.2em \raise 1.9ex\hbox{$...$}\kern -1.0em\hbox{$I$}}
\newcommand{\dddJ}{\kern 0.2em \raise 1.9ex\hbox{$...$}\kern-1.0em
\hbox{$J$}}
\newcommand{\dddcalJ}{\kern 0.2em \raise 1.9ex\hbox{$...$}\kern-1.0em
\hbox{${\cal J}$}}

\newcommand{\dddO}{\kern 0.2em \raise 1.9ex\hbox{$...$}\kern -1.0em
\hbox{${\cal O}$}}
\def\dddz{\raise 1.5ex\hbox{$...$}\kern -0.8em \hbox{$z$}}
\def\dddd{\raise 1.8ex\hbox{$...$}\kern -0.8em \hbox{$d$}}
\def\dddbd{\raise 1.8ex\hbox{$...$}\kern -0.8em \hbox{${\bf d}$}}
\def\ddbd{\raise 1.8ex\hbox{$..$}\kern -0.8em \hbox{${\bf d}$}}
\def\dddx{\raise 1.6ex\hbox{$...$}\kern -0.8em \hbox{$x$}}

\newcommand{\mplr}{m_{\rm Pl}}

\newcommand{\ode}{\Omega_{\rm DE}}
\newcommand{\oma}{\Omega_{M}}
\newcommand{\ora}{\Omega_{R}}

\newcommand{\rde}{\rho_{\rm DE}}

\newcommand{\pe}{\pa_{\eta}}

\newcommand{\DP}{\Delta_4}

\begin{document}

\begin{titlepage}

\vspace*{2cm}

\centerline{\Large \bf Conformal symmetry and  nonlinear extensions of nonlocal gravity}

\vskip 0.4cm
\vskip 0.7cm
\centerline{\large Giulia Cusin$^a$, Stefano Foffa$^a$, Michele Maggiore$^a$ and  Michele Mancarella$^{b,c,d}$}
\vskip 0.3cm
\centerline{\em $^a$D\'epartement de Physique Th\'eorique and Center for Astroparticle Physics,}  
\centerline{\em Universit\'e de Gen\`eve, 24 quai Ansermet, CH--1211 Gen\`eve 4, Switzerland}

\vspace{3mm}
\centerline{\em $^b$Institut de physique th\' eorique, Universit\'e  Paris Saclay}
\centerline{\em CEA, CNRS, 91191 Gif-sur-Yvette, France}
\vspace{3mm}
\centerline{\em $^c$Universit\'e Paris Sud, 15 rue George Cl\'emenceau, 91405,  Orsay, France}
\vspace{3mm}
\centerline{\em $^d$Physics Department, Theory Unit, CERN, CH-1211 Gen\`eve 23, Switzerland}

\vskip 1.9cm

\begin{abstract}
We study two nonlinear extensions of the nonlocal $R\,\Box^{-2}R$ gravity  theory. We extend this theory in two different ways suggested by conformal symmetry, either  replacing
$\Box^{-2}$ with $(-\Box + R/6)^{-2}$, which is the operator that enters the action for a conformally-coupled scalar field, or replacing $\Box^{-2}$  with the  inverse of the Paneitz operator,
which is  a four-derivative operator that enters in the effective action induced by the conformal anomaly.
We show that the former modification gives an interesting and viable cosmological model, with a dark energy equation of state today $w_{\rm DE}\simeq -1.01$, which  very closely mimics $\Lambda$CDM and  evolves  asymptotically into a de~Sitter solution. The model based on the Paneitz operator seems instead excluded by the comparison with observations. 
We also  review  some issues about the causality of nonlocal theories, and we point out that these nonlocal models can be modified so to nicely interpolate between Starobinski inflation in the primordial universe and accelerated expansion in the recent epoch.

\end{abstract}


\end{titlepage}

\newpage

\section{Introduction}

Much work has been recently devoted to the study of infrared  (IR) modifications of General Relativity (GR), with the aim of producing viable cosmological models displaying self-accelerating solutions even in the absence of a cosmological constant (see e.g. \cite{Bull:2015stt} for a recent review). In this context, our group has developed a program aiming at exploring the effect of nonlocal modifications of gravity. While at the fundamental level QFT is  local, at an effective level nonlocalities are commonly generated. This can happen both classically, when one integrates out some degrees of freedom to obtain an effective action for the remaining degrees of freedom, and at the quantum level, because of massless or light particles running into quantum loops. In principle this could generate nonlocal terms depending, e.g., on the inverse d'Alembertian $\iBox$. This operator becomes relevant in the IR, and is therefore potentially relevant in cosmology.

A nonlocal quantum effective action produces nonlocal equations of motion for the vacuum expectation values of the quantum fields. The relevant quantities, for cosmological applications, are the  in-in vacuum expectation values. The corresponding equations of motion, which are obtained using
the Schwinger-Keldysh formalism, depend on the inverse d'Alembertian defined with the  retarded Green's function, and are therefore automatically causal.\footnote{This point  has already been correctly discussed by several different groups, although there is still  some occasionally  confusion on it in the literature, and we take this opportunity to clarify the issue again in App.~\ref{sect:causality}.}

There are two aspects in the problem of developing a nonlocal IR modification of GR. First, at the purely phenomenological level, one must identify models which work well, i.e.  have a viable background evolution at the cosmological level, have well-behaved cosmological perturbations, and fit the observations, to the extent that they can compete with $\Lambda$CDM. Second, one must identify the specific mechanism that produces these nonlocalities from a fundamental local theory. It is quite natural to begin this program from the first part. Indeed, it is highly non-trivial to construct IR modifications of GR that are cosmologically viable, as has been learned  from experience with the DGP model \cite{Dvali:2000hr,Deffayet:2000uy,Deffayet:2001pu,Luty:2003vm,Nicolis:2004qq,Gorbunov:2005zk,Charmousis:2006pn,Izumi:2006ca}, the dRGT theory of massive gravity~\cite{deRham:2010ik,deRham:2010kj,Hassan:2011hr,Hinterbichler:2011tt,D'Amico:2011jj, DeFelice:2013awa,deRham:2014zqa}, 
Hassan-Rosen bigravity~\cite{Hassan:2011zd,Volkov:2011an,Comelli:2011zm,vonStrauss:2011mq,Akrami:2012vf,Comelli:2012db,Tamanini:2013xia,Fasiello:2013woa,Akrami:2013ffa,Konnig:2013gxa,Comelli:2014bqa,Solomon:2014dua,DeFelice:2014nja,Konnig:2014xva,Konnig:2014dna,Lagos:2014lca,Cusin:2014psa,Konnig:2015lfa,Akrami:2015qga,Schmidt-May:2015vnx}, or nonlocal models of the Deser-Woodard type~\cite{Deser:2007jk,Deser:2013uya,Woodard:2014iga,Dodelson:2013sma}. Indeed, none of these attempts has yet produced a viable competitor to $\Lambda$CDM. We can therefore hope that the condition of producing a cosmologically viable model will be sufficiently restrictive to select a limited range of nonlocal models. In turn, this might give precious hints for their derivation from a fundamental local theory.

The aim of this paper is to explore some possibly well-motivated nonlinear extensions of the nonlocal models that have been recently proposed by our group, in order to contribute to charting the  territory of possible viable nonlocal models. The paper is organized as follows. In Section~\ref{sect:over} we put our work into context, giving an overview of the different possibility that have been explored to date, and we will justify our choice of the class of models that deserve to be further investigated. In Sections~\ref{sect:nonmin} and ~\ref{sect:Paneitz} we will examine two particularly interesting nonlinear extensions of the simplest viable nonlocal model.
We present our conclusions in Section~\ref{sect:Concl}. We also take this opportunity to review, in 
App.~\ref{sect:causality}, some issues about the causality of nonlocal theories,  that occasionally generate some confusion. 
We use  the Misner, Thorne and Wheeler (MTW) conventions for the metric, Riemann tensor, etc., so in particular our  signature is $(-, +, +, +)$, and
we set $\hbar=c=1$. A prime will denote the derivative with respect to $x\equiv \log a$, where $a$ is the scale factor in FRW.

\section{An overview of nonlocal models}\label{sect:over}

The class of nonlocal models that we investigate here are characterized by the fact that the nonlocal terms are associated to an explicit mass scale $m$ (and are therefore different from the nonlocal models studied in \cite{Deser:2007jk,Deser:2013uya,Woodard:2014iga,Dodelson:2013sma} as well from those discussed in \cite{Barvinsky:2003kg,Barvinsky:2011hd,Barvinsky:2011rk}). The original inspiration came from the degravitation idea~~\cite{ArkaniHamed:2002fu,Dvali:2006su,Dvali:2007kt}, in which Einstein equations were modified phenomenologically into
\be\label{degrav}
\(1-\frac{m^2}{\Box}\)\Gmn=8\pi G\Tmn\, .
\ee
However, \eq{degrav} has the problem that the energy-momentum tensor is no longer automatically conserved, since in curved space the covariant derivative $\n_{\mu}$ does not commute with the covariant d'Alembertian $\Box$, and therefore does not commute with $\iBox$ either. One can however observe that any symmetric tensor $\Smn$ can be decomposed as
\be\label{splitSmn}
S_{\mu\nu}=S_{\mu\nu}^{\rm T}+\frac{1}{2}(\n_{\mu}S_{\nu}+\n_{\nu}S_{\mu})\, , 
\ee
where $S_{\mu\nu}^{\rm T}$ is the transverse part of the tensor, that satisfies
$\n^{\mu}S_{\mu\nu}^{\rm T}=0$. It can be proven that this decomposition is valid in a generic curved space-time~\cite{Deser:1967zzb,York:1974}. The extraction of the transverse part of a tensor is itself a nonlocal operation. For instance in flat space, where $\n_{\mu}\ra\pam$, applying to both sides of \eq{splitSmn} $\paM$ and  $\paM\paN$,
it is easy to show that the inversion of \eq{splitSmn} is
\be
S_{\mu\nu}^{\rm T}=\Smn
-\frac{1}{\Box}(\pam\paR S_{\rho\nu}+\pan\paR S_{\rho\mu})
+\frac{1}{\Box^2}\pam\pan\paR\paS S_{\rho\sigma}\, .
\ee
In a generic curved spacetime there is no such a simple formula. In any case, technically the easiest way to handle these models is to put them in a local form with the help of auxiliary fields, see below.
Using the possibility of extracting the transverse part of a tensor, in \cite{Jaccard:2013gla} it was proposed to modify \eq{degrav} into
\be\label{GmnT}
\Gmn -m^2\(\iBox\Gmn\)^{\rm T}=8\pi G\,\Tmn\, ,
\ee
so that energy-momentum conservation $\n^{\mu}\Tmn=0$ is automatically ensured. In \cite{Maggiore:2013mea,Foffa:2013vma} it was however found that the cosmological evolution that follows from this model is unstable, already at the background level. We will review below how such instabilities can in principle emerge in these nonlocal models. In any case, this adds the model (\ref{GmnT}) to the long list of IR modifications of GR that did not make it. 

The first successful nonlocal model was then proposed in \cite{Maggiore:2013mea}, observing that the instability is specific to the form of the $\iBox$ operator on a tensor such as $\Rmn$ or $\Gmn$, and does not appear when $\iBox$ is applied to a scalar, such as the Ricci scalar $R$. Thus, in  \cite{Maggiore:2013mea} it was proposed a model based on the nonlocal equation
\be\label{RT}
\Gmn -\frac{m^2}{3}\(\gmn\iBox R\)^{\rm T}=8\pi G\,\Tmn\, ,
\ee
where the factor $1/3$ provides a convenient normalization for the new mass parameter $m$. This model 
is quite interesting phenomenologically. It  has no van~Dam-Veltman-Zakharov  discontinuity, and smoothly reduces to GR in the limit $m\ra 0$. For $m={\cal O}(H_0)$, as will be required by cosmology, it therefore passes without difficulty all solar-system and laboratory constraints \cite{Maggiore:2013mea,Kehagias:2014sda}.\footnote{See also App.~A of \cite{Diraninprep} for the
clarification of an  issue on the comparison with Lunar Laser Ranging data. }
At the cosmological level, its background evolution is stable during RD and MD and has a self-accelerating solution, i.e. the nonlocal term behaves as  an effective  dark energy density~\cite{Maggiore:2013mea,Foffa:2013vma}. This produces  a realistic background FRW evolution, without the need of introducing a cosmological constant. Its cosmological perturbations are well-behaved, both  in the scalar  \cite{Dirian:2014ara} and in the tensor 
sector \cite{Cusin:2015rex}. The study of the effect of its cosmological perturbations shows that the predictions of the  model are  consistent with CMB, supernovae, BAO and structure formation 
data~\cite{Nesseris:2014mea,Dirian:2014ara,Barreira:2014kra}. The cosmological perturbations have then been implemented in a Boltzmann code in \cite{Dirian:2014bma}. This allowed us to perform Bayesian parameter estimation and a detailed quantitative comparison with $\Lambda$CDM, that shows that the model fits the data at a level which is statistically indistinguishable from  $\Lambda$CDM.\footnote{It should also be appreciated that this model only introduces one new parameter $m$, which replaces the cosmological constant in $\Lambda$CDM. By comparison, bigravity replaces the cosmological constant by a set of 5 parameters $\beta_n$, $n=0,\ldots , 4$ and also introduces a new Planck mass associated to the second metric, and viable solutions are searched tuning this parameter space. Similarly,
in the Deser-Woodard model one tunes  a whole function $f(\iBox R)$.} Having passed all these tests the model deserves a name, and 
we have dubbed it  the ``RT" model, where R stands for the Ricci scalar and T for the extraction of the transverse part.

A closed form for the action corresponding to \eq{RT} is currently not known. This model is however  closely related to another nonlocal model, subsequently proposed in \cite{Maggiore:2014sia}, and defined by the action
\be\label{RR}
S_{\rm RR}=\frac{\mplr^2}{2}\int d^{4}x \sqrt{-g}\, 
\[R-\mu R\frac{1}{\Box^2} R\]\, ,
\ee
where $\mplr$ is the reduced Planck mass, $\mplr^2=1/(8\pi G)$, and
$\mu\equiv m^2/6$. Indeed, if we compute the equations of motion  from \eq{RR} and we linearize them over Minkowski space, we  
find the same equations of motion  obtained by linearizing \eq{RT}. However, at the full nonlinear level, or linearizing over a background different from Minkowski, the two models are different. Also the model (\ref{RR}) works very well, both at the background level \cite{Maggiore:2014sia} and at the level of perturbations
\cite{Dirian:2014ara}. Again, the perturbations of this model have been implemented in a  Boltzmann code in\cite{Dirian:2014bma}, and compared to observations  using the  2013 Planck data. It was found that the model fits again well the data, even if not as well as $\Lambda$CDM or the RT model, although at the level of the analysis of
\cite{Dirian:2014bma} the difference was not statistically very significant. We will call the model defined by \eq{RR}  the ``RR" model. Further work on the RR and RT models has been presented 
in \cite{Modesto:2013jea,Foffa:2013sma,Conroy:2014eja,Cusin:2014zoa,Dirian:2014xoa,Mitsou:2015yfa,Barreira:2015fpa,Barreira:2015vra}.

Of course, as often happens in model building, there are  in principle infinite  choices for the specific form of the nonlocal model. We have therefore attempted to chart this large unexplored territory, considering some particularly natural extensions of these models. At the level of models defined by the equations of motion using the extraction of the transverse part, we have seen that the RT model (\ref{RT}), where $(\gmn \iBox R)^T$ enters,  is viable, while a model where appears   $(\iBox\Gmn)^T$, or equivalently where appears  $(\iBox\Rmn)^T$, is not. These models are naturally written down at the level of equations of motions, but are not easily written in terms of actions. 
Turning to models defined at the level of the action, one can observe that
a basis for the curvature-square terms is provided by $\Rmnrs^2$, $\Rmn^2$ and $R^2$. However, for cosmological applications it is more convenient to trade the Riemann tensor $\Rmnrs$ for the Weyl tensor $C_{\rho\sigma\mu\nu}$. Thus,
a natural generalization of the RR model is given by
\be\label{actionTotal}
S=\frac{\mplr^2}{2}\int d^4 x \sqrt{-g}\,
\left[R-\mu_1 R\frac{1}{\Box^2}R-\mu_2 C^{\mu\nu\rho\sigma}\frac{1}{\Box^2}C_{\mu\nu\rho\sigma}-\mu_3\RMN\frac{1}{\Box^2}\Rmn
\right]\,,
\ee
where $\mu_1$, $\mu_2$ and $\mu_3$ are  independent parameters with dimension of squared mass.
In \cite{Cusin:2015rex} is has however been found that the term $\RMN\Box^{-2}\Rmn$ is ruled out since it gives again instabilities at the background level.\footnote{This result is analogous to the one found in \cite{Ferreira:2013tqn}, where it was shown that a term 
$\RMN\Box^{-1}\Rmn$ also produces instabilities in the cosmological evolution. Observe that the latter term is rather of the Deser-Woodard type, i.e. of the form $\RMN f(\Box^{-1}\Rmn)$, with a dimensionless function $f$ and no explicit mass scale $m$. However, in both cases the instability is ultimately due to the form of the $\iBox$ operator on the tensor $\Rmn$, as also in the model (\ref{GmnT}).} The Weyl-square term instead does not contribute to the background evolution, since the Weyl tensor vanishes in FRW, and it also has well-behaved scalar perturbations. However, its tensor perturbations are unstable \cite{Cusin:2015rex}, which again rules out this term.\footnote{Unless one takes a model for the early Universe that generates a totally negligible amount of primordial tensor perturbations, so there are no primordial tensor perturbations that will be amplified by the subsequent unstable evolution. Within the standard inflationary paradigm, even for models with extremely low values of the tensor-to-scalar ratio, the Weyl term is ruled out.} 

These results show that the condition of obtaining a viable cosmological model is indeed a powerful requirement, which allows us to eliminate most of the possible choices. In practice, at least within the space of theories that we have explored, we find that only models constructed uniquely with the Ricci scalar work. At a finer level of resolution, using again the  Boltzmann code modified for nonlocal theories, in \cite{Diraninprep} we have repeated the comparison  with observations   using the 2015 Planck data (which were not yet publicly available when \cite{Dirian:2014bma} appeared) as well as with an extended set of BAO and structure formation data. A Bayesian model comparison between $\Lambda$CDM, the RT and RR models has then been performed. In this improved analysis, $\Lambda$CDM and the RT models still both fit the data very well, and are statistically indistinguishable. In contrast the RR  model, while by itself still fits the data at a fully acceptable level, in a Bayesian model comparison with $\Lambda$CDM or with the RT model is now significantly disfavored.

In a sense, the RT model can be considered as a nonlinear extension of the RR model, since the two models become the same when linearized over Minkowski. An action for the RT model would probably include further nonlinear terms, such as higher powers of the curvature associated to higher powers of $\iBox$. Since the data seem to point toward the importance of these nonlinear terms, it is natural to ask whether other nonlinear extensions of the RR theory are cosmologically viable. Once again, it is not possible  to explore the most general form of these extensions. 
However, symmetries are often a powerful guide for model building. In particular, conformal symmetry naturally appears at high energies, or in the presence of massless particles.
In the physical and mathematical literature, there are two notable extensions of the $\Box$ or of the $\Box^2$ operator, related to conformal symmetry. The first is obtained replacing 
\be\label{Boxxi}
-\Box\ra -\Box +\frac{1}{6} R\, .
\ee
This is the operator that appears in the action of a conformally coupled scalar field  in $D=4$ space-time dimensions. Another interesting option is to replace directly $\Box^2$ with the Paneitz operator
\be\label{defDP}
\DP\equiv\Box^2+2\RMN\n_{\mu}\n_{\nu}-\frac{2}{3}R\Box+\frac{1}{3}\gMN\n_{\mu} R\n_{\nu}\, .
\ee
This operator was independently discovered  in a mathematical context, as well as in physics in the context of conformal supergravity~\cite{Fradkin:1982xc}, and there is a large body of mathematical literature on it. In physics  $\Delta^{-1}_4$  appears 
in particular in the nonlocal anomaly-induced effective action in four dimensions (see e.g. \cite{Shapiro:2008sf} for review, and \eq{SanomD4} below).
Just as the operator (\ref{Boxxi}), $\Delta_4$ only depends on the conformal structure of the space-time.\footnote{The Paneitz operator was also considered in the context of the Deser-Woodard class of nonlocal models  in \cite{Deser:2007jk}, where was considered the possibility of adding to the Ricci scalar in the action a term $R\Delta_4^{-1}R^2$ which, on dimensional ground, does not require the introduction of a mass scale.}

The first nonlinear extension of the RR model that we will consider is then defined by the action
\be\label{6RR}
S_{\rm cRR}=\frac{\mplr^2}{2}\int d^4x\, \sqrt{-g}\, 
\[ R-\mu R\frac{1}{(-\Box+  \frac{1}{6}R)(-\Box +\frac{1}{6} R)} R\]\, .
\ee
We will call it the ``conformal RR" model.
The second model that we will investigate is defined by
\be\label{actDelta4}
S_{\Delta_4}=\frac{\mplr^2}{2}\int d^4x\, \sqrt{-g}\, \[ R-\mu R\frac{1}{\DP} R\]\, .
\ee
In both case,  $\mu\equiv m^2/6$. If the model (\ref{6RR}) should work, this could give a hint that the fundamental theory behind the effective nonlocal model involves a conformally-coupled scalar field. The model
(\ref{actDelta4}) would rather point toward a role of the anomaly-induced effective action in the derivation of the nonlocal theory, possibly along the lines recently discussed in \cite{Maggiore:2015rma}.

To assess whether a model is cosmologically viable we will study its cosmology at the background level, verifying if it has a stable self-accelerating background solution. If this is the case, one should then in principle study its cosmological perturbations, implement them in a Boltzmann code, perform parameter estimation of the model,  and finally compare with the data. The latter part is of course very laborious. However, an approximate but much simpler criterium for the viability of the model is given by its prediction for the equation of state (EOS) of DE. Dynamical DE models are often investigated using the so-called $w$CDM model, in which the EOS parameter is taken to be a constant in time, with a value $w$ which, rather than being fixed to $-1$ as in $\Lambda$CDM, is taken as a new fitting parameter. Actually, once a dynamical DE is considered, there is no reason a priori why $w$ should be constant in time, and a more general phenomenological parametrization is obtained in the so-called  $(w_0,w_a)$ model, where near the recent epoch $w(a)$ is written, as a function of the scale factor $a$, as 
\be\label{w0wa} 
w(a)=w_0+(1-a)w_a\, .
\ee
Of course, both in $w$CDM  and in the $(w_0,w_a)$ parametrization no DE perturbations are included, so these parametrizations do not 
exactly capture all features of a specific dynamical model, such as the RR and RT nonlocal models or of their generalizations. Indeed, these specific nonlocal  models  also have a given structure of cosmological perturbations, which differs $\Lambda$CDM, and which also affect the parameter estimation in these models. Thus, to perform a quantitative Bayesian comparison between the performance of a nonlocal model with that of $\Lambda$CDM, there is no alternative to the full analysis, as done for the RT and RR models in  \cite{Dirian:2014bma,Diraninprep}. However, to have a first estimate of whether a model is viable, we can just compare the value of $w(a)$ obtained from the nonlocal model with the limits on $w_0$ or on $(w_0,w_a)$ obtained comparing $w$CDM or the $(w_0,w_a)$ model to the data, which has been done in the 2015 Planck dark energy paper \cite{Ade:2015rim}. Indeed, experience with the RT and RR model shows that this criterium gives quite reasonable results. In particular, for the RT model, one finds \cite{Maggiore:2013mea}
$w_0\simeq -1.04$, $w_a\simeq -0.02$,\footnote{Of course the result depends on the value chosen for the matter density fraction $\oma$, which in \cite{Maggiore:2013mea} was fixed to the best-fit value obtained by Planck from $\Lambda$CDM. In the full analysis including cosmological perturbations, $\oma$ is eventually determined  by the Bayesian parameter estimation. However, $w_0$ has a  weak dependence on the precise value of $\oma$, which also does not change much between  $\Lambda$CDM and the nonlocal model. Then, the full analysis confirms this value of $w_0$, at the level $\Delta w_0=0.01$.} while for the RR model $w_0 \simeq -1.14$, $w_a = 0.08$
\cite{Maggiore:2014sia}. This suggests that the RT model produces deviations, with respect to $\Lambda$CDM, of order of a few percent, while the RR model should produce larger deviations. Indeed,
in the full Boltzmann code analysis  we found for instance that, in structure formation, the RT model shows deviation from $\Lambda$CDM at the level of about $2\%$, while the RR model shows deviations that, depending on the observable, can be up to $8\%$ \cite{Dirian:2014bma,Diraninprep}. Similarly, when performing parameter estimation from CMB, SNe and BAO, the results for the
RT model are quite close to that of $\Lambda$CDM, while the RR model shows larger departures from
$\Lambda$CDM. For instance, the best-fit values for $H_0$  from Planck 2015 temperature and polarization data, plus the set of BAO and SNa data considered in \cite{Diraninprep} are
$H_0=67.67$ for $\Lambda$CDM, $H_0=68.76$ for the RT model and $H_0=70.44$ for the RR model.

In  \cite{Diraninprep}, performing the  Bayesian comparison between the model, we found that  the RR model, which has  $w_0\simeq -1.14$,  is disfavored, while the RT model, with a value $w_0\simeq -1.04$ closer to the $\Lambda$CDM value $w=-1$, is fully consistent with the observations, and fit the data in a way statistically equivalent to $\Lambda$CDM. These results are fully consistent with those obtained in the 2015 Planck dark energy paper \cite{Ade:2015rim} for the generic $w$CDM or  $(w_0,w_a)$ parametrizations.
This gives us a first guidance into the typical values that of $w_0$ that a nonlocal model should have, to be consistent with the observation. Of course, for a model that passes this first test, in the end a full analysis will be necessary, especially if we want to compare its performances to that of  $\Lambda$CDM.

\section{The conformal RR model}\label{sect:nonmin}

We first consider the model defined by 
\be\label{xiRR}
S_{\rm \xi RR}=\frac{\mplr^2}{2}\int d^4x\, \sqrt{-g}\, \[ R-\mu R\frac{1}{(-\Box+  \xi R)(-\Box +\xi R)} R\]\, .
\ee
Writing the action in terms of a generic value of $\xi$ can be useful at the mathematical level, to investigate the dependence on $\xi$. However, physically, beside $\xi=0$ there is only  one special value, which is
the conformal case $\xi=1/6$. To compare with the observational data we will only be interested  in the case $\xi=1/6$, which gives a sharp and physically-motivated prediction. Furthermore, if we keep $\xi$ as a free parameter and consider $\xi\neq 1/6$, we are no longer protected by conformal symmetry, and nothing forbids to add to $-\Box$ also a mass term, 
which would lead to a second extra free parameter. In this sense, the model (\ref{xiRR}) with $\xi=1/6$ is privileged also with respect to the RR model, which has $\xi=0$, or the RT model.
The model (\ref{xiRR}), which is a natural extension of the model (\ref{RR}), was already   studied in 
\cite{Mitsou:2015yfa}, closely following the  analysis of the RT and RR models in \cite{Maggiore:2013mea,Maggiore:2014sia}. Since however  the most interesting case $\xi=1/6$ was not specifically investigated in \cite{Mitsou:2015yfa}, we will repeat below part of this analysis, and the relevant numerical integration, and we will work out the prediction for the DE equation of state in this case.

Following a technique introduced in \cite{Nojiri:2010pw} in the context of the Deser-Woodard model,
and already used in \cite{Maggiore:2013mea,Maggiore:2014sia},
we write the action in a local form introducing two Lagrange multipliers $\lambda_1$ and $\lambda_2$ and two auxiliary scalar fields $S,U$, as 
\be\label{actconfLapl2}
S_{\xi RR}=\frac{\mplr^2}{2}\int d^4x\, \sqrt{-g}\, \big\{ R(1-\mu  S)
+\lambda_1\[ (-\Box +\xi R)U-R\] +\lambda_2\[ (-\Box +\xi R)S-U\]\big\} \, .
\ee
The variation with respect to the Lagrange multipliers enforces the equations
\bees
&&(-\Box +\xi R)U=R\, ,\label{BoxU}\\
&&(-\Box +\xi R)S=U\, ,\label{BoxS}
\ees
so in particular
\be
S=\frac{1}{(-\Box +\xi R)} U=\frac{1}{(-\Box +\xi R)(-\Box +\xi R)} R\, .
\ee
The variations with respect to $S$ and $U$ give $\lambda_1=\mu S$ and $\lambda_2=\mu U$.
The equations of motions obtained performing the variation with respect to the metric give (adding also the matter action)
\be\label{Gmn}
\Gmn=\mu K_{\mu\nu}+8\pi G\Tmn\, ,
\ee
where
\bees
K_{\mu\nu}&=&2(S-\xi US)\Gmn-2\n_{\mu}\pan S +2\gmn\Box (S-\xi US)+\gmn\parho S\paR U\nn\\
&& -(1/2)\gmn U^2-(\pam S\pan U+\pan S\pam U)\, ,
\label{K}
\ees
in agreement with \cite{Mitsou:2015yfa}.
For $\xi=0$, \eqss{BoxU}{BoxS}{K}  reduce to  that given in \cite{Maggiore:2014sia}. 

\subsection{Cosmological equations}

We now specialize to a FRW metric. The computation is a straightforward generalization of that performed in  \cite{Maggiore:2014sia}.  
We parametrize the time evolution using the variable $x=\log a$,  we denote $df/df=f'$ and
we introduce the notations
\be\label{4defV}
V(x)=H_0^2 S(x)\, ,\qquad \gamma =\frac{m^2}{9H_0^2}\, ,
\ee
as well as $h(x)=H(x)/H_0$, $\zeta(x)=h'(x)/h(x)$ and  
$\Omega(x)=\rho(x)/\rho_0$, where $\rho_0=3H_0^2/(8\pi G)$ is the critical density. We write 
\be
\Omega(x)=\oma e^{-3x}+\ora e^{-4x}\, , 
\ee
where $\oma$ and $\ora$ are the density fractions of matter and radiation today, respectively. From the $(00)$ component of \eq{K} we get the Friedman equation which,  in these dimensionless variables, reads
\be\label{Lfun1}
h^2(x)=\frac{\Omega(x) +(\gamma/4) U^2}{ 1+\gamma[ -3(V-\xi UV)'-3 (V-\xi UV) +(1/2)V'U' ]}\, ,
\ee
while \eqs{BoxU}{BoxS}  become
\bees
&&U''+(3+\zeta)U'+6\xi (2+\zeta) U=6(2+\zeta)\, .\label{Lfun2}\\
&&V''+(3+\zeta)V'+6\xi (2+\zeta)V=h^{-2}U\, ,\label{Lfun3}
\ees
This form is the most convenient for the numerical integration. For studying the stability of the system and for identifying the effective dark energy, it is convenient to trade $V$ for $W=H^2S=h^2 V$. Then
\be\label{h2OY}
h^2(x)=\Omega(x)+\gamma Y\, ,
\ee
where
\be\label{LdefY}
Y= \frac{1}{2}W'(6-U'-6\xi U) +W (3-6\zeta+\zeta U'-3\xi U'+6\xi\zeta U)+\frac{1}{4}U^2-3\xi UV\, .
\ee
\Eq{h2OY} shows that there is an effective dark energy term, with $\rde=\rho_0\gamma Y$, so the DE density
fraction is
\be\label{defode}
\ode (x)=\gamma Y\, .
\ee
\Eq{BoxS} becomes
\be\label{syW}
W''+3(1-\zeta) W'-2[\zeta'+3\zeta-\zeta^2-3\xi(2+\zeta)]W= U\, .
\ee
The fundamental equations for the variables $h(x)$, $U(x)$ and $W(x)$ are \eqsss{h2OY}{LdefY}{Lfun2}{syW}.

\subsection{Stability of the background solution}

An important aspect for the viability of a cosmological model is the stability of the background solution. A full analysis of the stability requires the study of linear cosmological perturbations, and we will report on it in a future work. However, already at the background level that we are considering in this paper, a stringent test is possible~\cite{Maggiore:2013mea,Foffa:2013vma,Maggiore:2014sia}. Indeed, the auxiliary fields $U$ and $W$ obey the inhomogeneous differential equations (\ref{Lfun2}) and (\ref{syW}). The general solution will be a superposition of a particular solution of the inhomogeneous equation and the most general solution of the associated homogeneous equations.
The latter can be easily obtained analytically whenever we are deep in a given era, so that $\zeta(x)$ becomes approximately a constant $\zeta_0$. In particular  $\zeta_0=\{-2,-3/2,0\}$ in RD, MD and de~Sitter (dS), respectively.
Then the solutions of the homogeneous equations associated to \eqs{Lfun2}{syW} have the general form 
$U_{\rm hom}=e^{\alpha_{\pm} x}$ and $W_{\rm hom}=e^{\beta_{\pm} x}$. If at least one among the four coefficients  $\alpha_{\pm}$, $\beta_{\pm}$ is positive, either in RD or in MD, there will be at least one growing mode. Of course, at the background level, one can in principle choose initial conditions such that 
$U_{\rm hom}=W_{\rm hom}=0$. However this is a fine-tuning, and any  spatially-homogeneous perturbation
$\d U(t), \d W(t)$
will move the system away from this point. Then, for a generic perturbation the growing mode will unavoidably be excited, and the background solution will be  destabilized. In other words, if the homogenous equations for $U$ and $W$ have growing modes, we automatically know that, when we will study linear cosmological perturbations, the variables $\d U(\vk, t)$ and $\d W(\vk, t)$ will show instability already in the spatially-homogeneous limit $\vk\ra 0$. The absence of growing modes for the homogeneous solutions is therefore a necessary (but certainly in general not sufficient) condition for the stability at the level of linear cosmological perturbations, and the homogeneous solutions must be stable both in RD and in MD.\footnote{In principle stability in the dS era is not mandatory because one might imagine that, at the large energy scales corresponding to primordial inflation, the nonlocal models are modified. Indeed, one could imagine that effective actions such as (\ref{RR}) might be valid only in the low energy limit, and could be modified at the large energies corresponding to inflationary scales. We will give an interesting example of this sort in \eq{6RRStar} below. In the numerical solution of the equations, we always start the integration deep in RD.
However, if the model is already stable even in de~Sitter, this is certainly a positive feature. Observe that  the RT model is only stable in RD and MD \cite{Maggiore:2013mea,Foffa:2013vma}, while the RR model is stable in dS, RD and MD \cite{Maggiore:2014sia}. We will find below that also  the model (\ref{xiRR}) with $\xi=1/6$ is stable in all three eras, see also \cite{Mitsou:2015yfa} for $\xi$ generic.}

It is indeed this stability criterium that ruled out the model (\ref{GmnT}), while the RR and RT models passed this test
\cite{Maggiore:2013mea,Foffa:2013vma,Maggiore:2014sia}. For the model (\ref{xiRR}), specializing directly to the physically interesting case $\xi=1/6$, the homogeneous equation for $U$, with $\zeta=\zeta_0$ constant, reads
\be\label{fun3hom}
U''+(3+\zeta_0)U+ (2+\zeta_0) U=0\, .
\ee
The corresponding solutions are $U=e^{\a_{\pm} x}$ with $\a_{+}=-1$ and $\a_{-} =-(2+\zeta_0)$, which are never positive in  RD, MD or de~Sitter. Similarly, the solutions of the homogeneous equation for $W$ are
$W=e^{\b_{\pm} x}$ with $\b_{+}=-1+2\zeta_0$ and $\b_{-} =-2+\zeta_0$, which again are both negative, in all three eras. Therefore, there is no instability in the background evolution, as also observed in \cite{Mitsou:2015yfa}. 

\subsection{Solution for the background evolution}

For the numerical integration we use 
\eqst{Lfun1}{Lfun3}. As  in  \cite{Dirian:2014ara}, we observe that
in \eqs{Lfun2}{Lfun3} appears $\zeta=h'/h$. However, $h'$ can be computed explicitly taking the derivative of
\eq{Lfun1}. The resulting expression contains $V''$ and $U''$, which  can  be eliminated using again
\eqs{Lfun2}{Lfun3}.
The result is  given by 
\be
\zeta= \frac{h^{-2}\Omega'+3\g Z}{2[1-3 \gamma V (1-6\xi)(1-\xi U)]}
\, .
\ee
where
\be
Z=h^{-2}U+U'V'-4V'-\xi (  24V-4UV'-4U'V+2U'V'+h^{-2}U^2) +24\xi^2UV\, .
\ee
The value of $\gamma$ is tuned so to obtain the desired value of $\oma$ (i.e., choosing a value of $\oma$ and tuning $\gamma$ requiring that the solution of the numerical integration satisfies $h(x=0)=1$, which follows from the definition $h(x)=H(x)/H_0$). Setting $\oma=0.313$ we get
$\gamma\simeq 0.081$ and therefore 
\be\label{m085}
m\simeq 0.85 H_0\, .
\ee
Finally, from \eqs{defY}{defode}, the DE fraction is obtained 
by $\ode(x)=h^2(x)-\Omega(x)$. From this we can then get the DE equation of state parameter $w_{\rm DE}$, defined as usual by 
\be\label{consrho}
{\rho}'_{\rm DE}+3(1+w_{\rm DE})\rho_{\rm DE}=0\, .
\ee
(recall that the prime is the derivative with respect  to $x$), so

\begin{figure}[t]
\centering
\includegraphics[width=0.48\columnwidth]{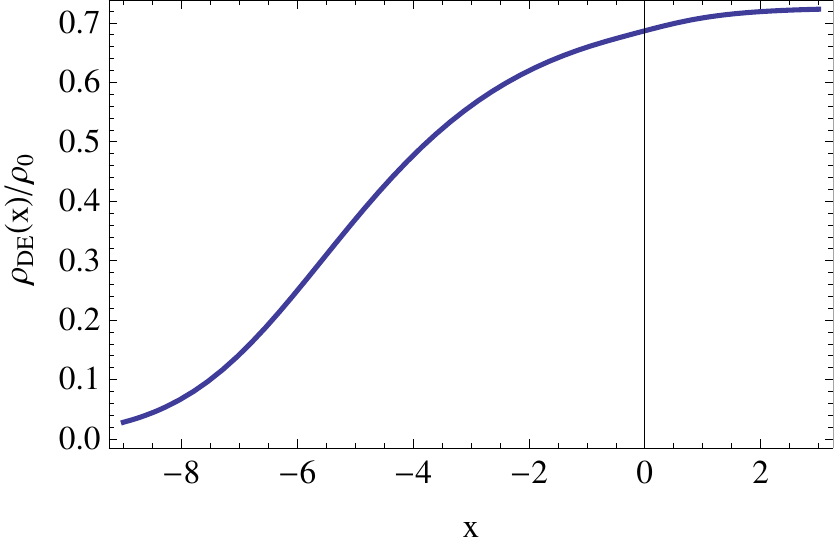}
\includegraphics[width=0.48\columnwidth]{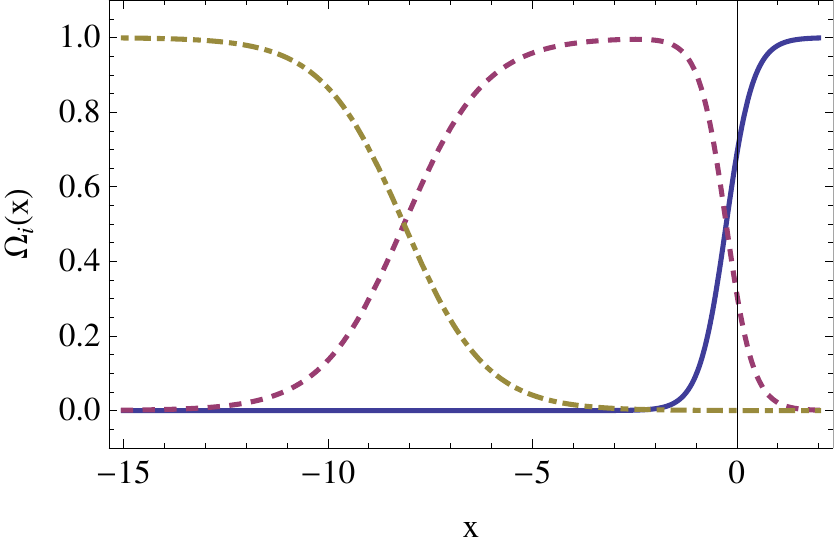}
\includegraphics[width=0.48\columnwidth]{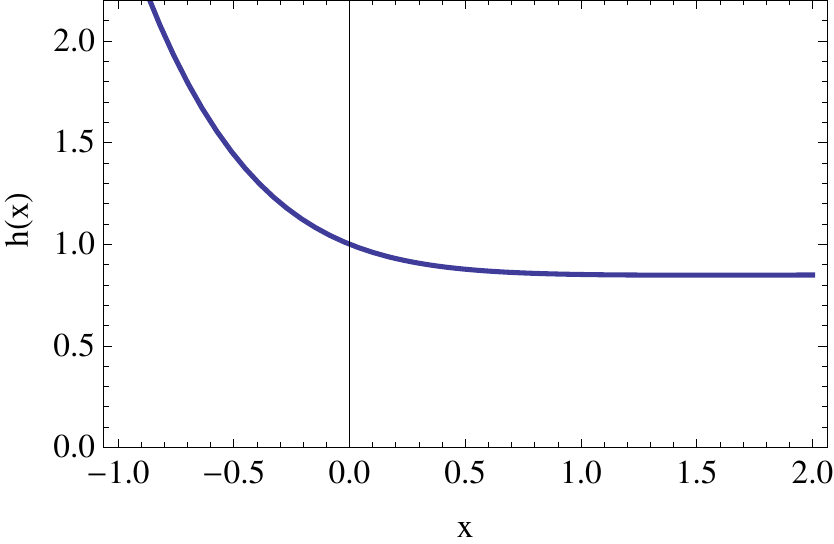}
\caption{\label{Lfig:rho} 
Top left panel: the dark energy density $\rde(x)/\rho_0$.
Top right panel: the energy fractions of radiation (dot-dashed), matter (dashed) and dark energy (solid line).
Bottom panel:  the Hubble parameter $h(x)$.
}
\end{figure}
\begin{figure}[t]
\centering
\includegraphics[width=0.48\columnwidth]{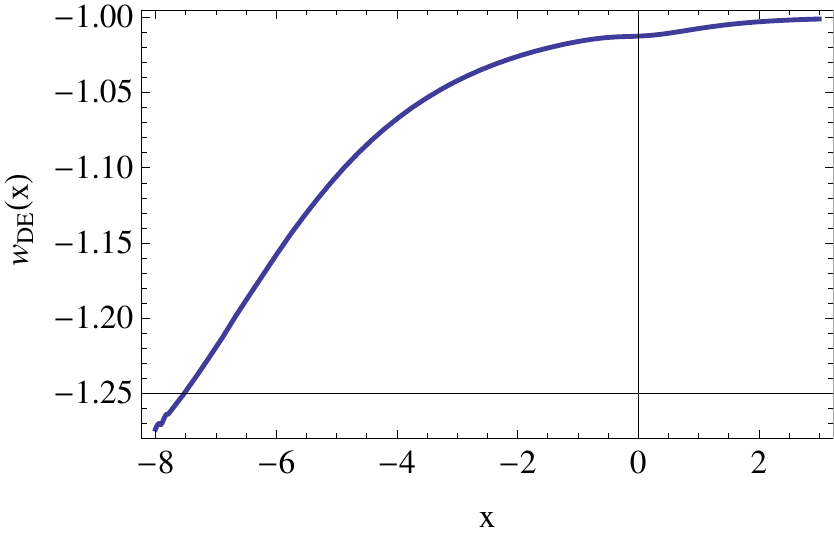}
\includegraphics[width=0.48\columnwidth]{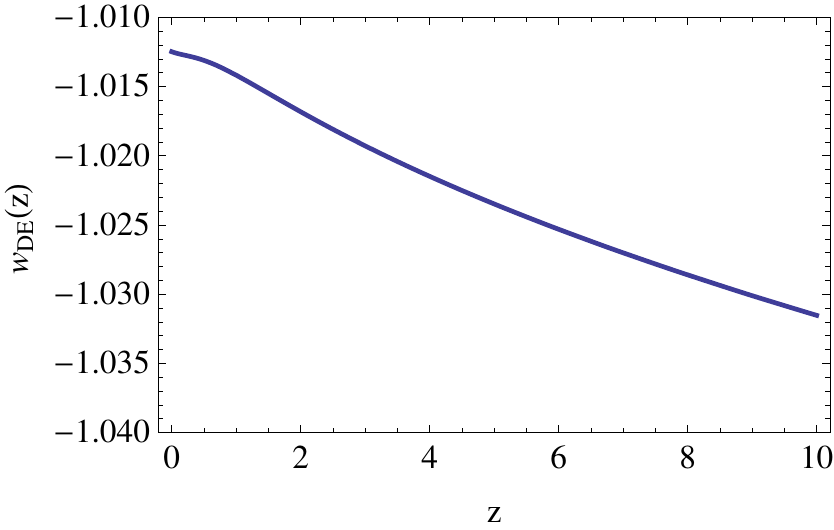}
\includegraphics[width=0.48\columnwidth]{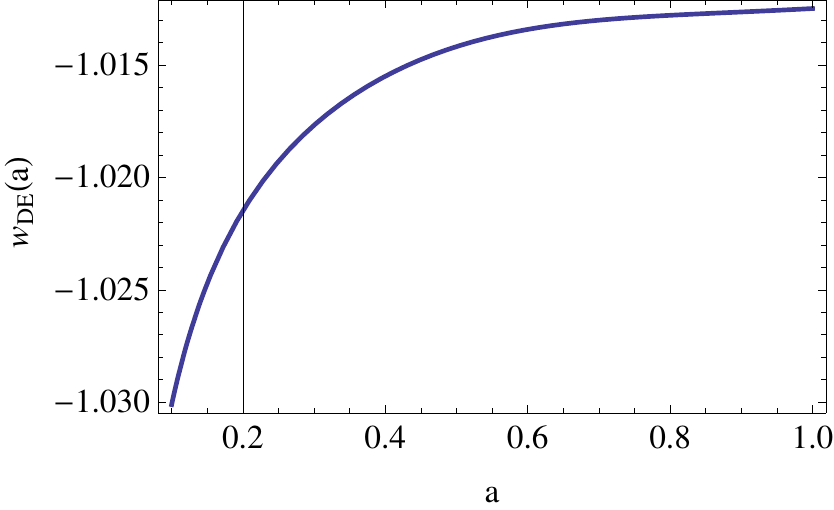}
\caption{\label{Lfig:wDE} 
The EOS parameter of dark energy, $w_{\rm DE}$, as a function of $x=\ln a$ (top left), of redshift $z$ (top right) and  of scale factor $a$ (bottom). }
\end{figure}


\be
w_{\rm DE}=-1-\frac{{\rho}'_{\rm DE}}{3{\rho}_{\rm DE}}=-1-\frac{\ode'}{3\ode}\, .
\ee
The results of the numerical integration  are shown in Figs.~\ref{Lfig:rho} and \ref{Lfig:wDE}. We see from Fig.~\ref{Lfig:rho} that, asymptotically, the Hubble parameter and the DE density go to a constant. The model therefore goes asymptotically to de~Sitter, just as  $\Lambda$CDM, as also observed in~\cite{Mitsou:2015yfa}. This result is easily understood analytically.
Rewriting \eqst{Lfun1}{Lfun3} in terms of cosmic time $t$ and of the Hubble parameter $H(t)=\dot{a}/a$,  we get
\be\label{H2cosmict}
H^2=\frac{m^2}{9}\[ 3H\pa_t(S-\xi US)+3H^2(S-\xi US)-\frac{1}{2}\dot{S}\dot{U}+\frac{1}{4}U^2\]
+\frac{8\pi G}{3}\rho\, ,
\ee
\bees
&&\ddot{U}+3H\dot{U}+12\xi^2H^2U=12H^2\, ,\label{Ucosmict}\\
&&\ddot{S}+3H\dot{S}+12\xi^2H^2S=U\, ,\label{Scosmict}
\ees
where $\dot{f}\equiv\pa f/\pa t$. We now consider the asymptotic regime where DE dominates (see the top-right panel of Fig.~\ref{Lfig:rho}), so the term $(8\pi G/3)\rho$ on the right-hand side of \eq{H2cosmict} is negligible compared to the effective DE term, and in this limit we look for a solution with $H$ constant. In this case, for $\xi\neq 0$, the solution of \eq{Ucosmict} is
\be\label{U1suxi}
U=\frac{1}{\xi}+u_1 e^{c_1 Ht} +u_{2} e^{c_2 Ht} \, ,
\ee
where $2c_{1,2}=-3\pm\sqrt{9-48\xi}$. The real part of $c_1$ and $c_2$ is always negative, for all positive values of $\xi$. In particular, for $\xi=1/6$ we have $c_1=-1$ and $c_2=-2$. Therefore the solutions of the homogeneous equation decay exponentially in time and, asymptotically, $U\ra 1/\xi$. Similarly, $S\ra 1/(2\xi H^2)$, plus the same exponentially decaying solutions. Plugging these constant values for $U$ and $S$ in \eq{H2cosmict} and neglecting the term proportional to $\rho$ we see that $H= {\rm constant}$ is indeed a solution, and we find
\be\label{H}
H=\frac{m}{6\xi}\, .
\ee
In particular, for $\xi=1/6$, asymptotically $H= m$, and therefore $h= 3\gamma^{1/2}$. This perfectly agrees with the asymptotic value of $h$ that we find from the numerical integration, shown in the bottom panel of Fig.~\ref{Lfig:rho}. Observe that the case $\xi\neq 0$ is quite different from the RR model, which corresponds to $\xi=0$ and therefore does not have this solution. Indeed, in the RR model $h(x)$ grows indefinitely, even in the DE dominated era, see Fig.~1 of ref.~\cite{Maggiore:2014sia}.

Alternatively, we can understand the emergence of a de~Sitter solution directly from the action, observing that in a regime of constant (and non-vanishing) $R$, the operator $(-\Box+\xi R)^{-1}$ acting on $R$ reduces to $(\xi R)^{-1}$. Then the nonlocal term in the action (\ref{xiRR}) reduces to a cosmological constant $\Lambda=m^2/(12\xi^2)$, leading  to 
a de~Sitter era with $H^2=(1/3)\Lambda$, in agreement with \eq{H}.

The DE equation of state is shown in Fig.~\ref{Lfig:wDE}, and we see that
it is  on the phantom side, as for the RR and RT models. This  is a  consequence of \eq{consrho}, together with the fact that in these three models the DE density vanishes in RD and then grows monotonically, so
${\rho}'_{\rm DE}>0$ and $\rho_{\rm DE}>0$, which implies $1+w_{\rm DE}<0$. Among the three plots in
Fig.~\ref{Lfig:wDE}, the one against $x$ shows $w_{\rm DE}$ on the largest time interval. The matter-radiation equilibrium is around $x\simeq -8.1$, so this plot displays the whole MD era, the present DE dominated era and is extended into the future, $x>0$. The plot as a function of redshift $z$ focus on the more recent past epoch  $0\leq z\, \lsim \, 10$. This is still  a range broader  than that on which such a DE  can play a relevant role (observe that, going backward in time, this DE decreases, rather than being a constant as in $\Lambda$CDM, so in the past it becomes even  more and more irrelevant than a cosmological constant). Note that $z=10$ corresponds to $x\simeq -2.4$. Observing that the vertical scale has been expanded, we see that the dependence of $w_{\rm DE}$ on $z$ in this epoch is very mild, and $w_{\rm DE}$ is very close to $-1$. Indeed, $w_{\rm DE}(z=0)\simeq -1.012$. 
The bottom panel in Fig.~\ref{Lfig:wDE} shows that, in the recent epoch, the standard linear fit (\ref{w0wa}) is appropriate. Fitting our numerical result to  \eq{w0wa} for $-1<x<0$, we get
\be\label{Lfitw0wa}
w_0\simeq -1.012\, ,\qquad w_a=-0.005\, .
\ee 
Actually, Fig.~\ref{Lfig:wDE} shows that a fit linear in $z$ is appropriate on a broader range, so we also perform a fit of the form
\be
w(z)=w_{0}+w_{z} z\, .
\ee
In this case, the fit is excellent in the whole  range $0<z<10$ (which corresponds to $-2.4<x<0$), and we get
\be\label{Lfitw0wz}
w_0\simeq -1.012\, ,\qquad w_z=-0.002\, .
\ee 
Since also the perturbations in the dark energy sector are proportional to $1+w_{\rm DE}$, we expect  that this model produces deviations from $\Lambda$CDM at the level of about $1\%$, which are not detectable with present observations. A  more detailed quantitative analysis will require the computation of the cosmological perturbations and their implementations in a Boltzmann code, as performed in \cite{Dirian:2014ara,Dirian:2014bma,Diraninprep} for the RR and RT nonlocal model. We hope to report on this in the future. However, from the discussion at the end of Section~\ref{sect:over}, it is clear that the model is consistent with observations, and we expect that its predictions will be intermediate between that of $\Lambda$CDM and that of the RT model.

\section{Non-local action with the Paneitz operator}\label{sect:Paneitz}

We next consider the action (\ref{actDelta4}).
The operator $\Delta_4$ has a number of interesting mathematical properties. In particular, if two metrics $\gmn$ and $\gbmn$ are related by a conformal factor, $\gmn=e^{\phi}\gbmn$, then
\be\label{barDP}
\sqrt{-g}\,\DP=\sqrt{-\bar{g}}\, \bar{\Delta}_4\, .
\ee
In this sense, it is the analogous of the Laplacian in two dimensions, and indeed it also appears in the four-dimensional quantum effective action for the conformal anomaly. We rewrite $S_{\Delta_4}$ introducing an auxiliary field $S$ and a Lagrange multiplier $\xi$,
\be
S_{\Delta_4}=\frac{\mplr^2}{2}\int d^4x\, \sqrt{-g}\, \[ R-\mu R S-\xi (\DP S-R)\]\, .
\ee
The variation with respect to $\xi$ enforces 
\be\label{DPSR}
\DP S=R\, ,
\ee
i.e. $S=\DP^{-1}R$.  In covariant form, the equations of motions are
\bees
&&G_{\alpha\beta}\left[1-\mu S + \xi - \frac{2}{3}\left(\n_\rho\xi\right)\n^\rho S\right]+g_{\alpha\beta}\left[\Box \xi -\mu \Box S-\frac{4}{3} R^{\rho\sigma}\left(\n_\sigma\xi\right)\n_\rho S\right.\nonumber\\
&&-\left.\frac{1}{6}\left(\n_\rho\Box\xi\right)\n^\rho S-\frac{1}{6}\left(\n_\rho\xi\right)\n^\rho \Box S+\frac{1}{2}\left(\Box\xi\right)\Box S-\frac{1}{3}\left(\n_\rho \n_\sigma\xi\right)\n^\rho \n^\sigma S\right]\nonumber\\
&&+\n_\alpha\n_\beta\left[\mu S -\xi-\xi \Box S -\frac{1}{3}\left(\n^\rho\xi\right)
\n_\rho S\right]
-\frac{2}{3}R\left(\n_{(\alpha}\xi\right)\n_{\beta)} S
-\xi\left(\n^\rho R_{\alpha\beta}\right)\n_\rho S\nonumber\\
&&+2\left(\n^{\rho}\xi\right)R_{\rho(\alpha}\n_{\beta)} S
+2\left(\n_{(\alpha}\xi\right)R_{\beta)\rho}\n^\rho S
-\Box\xi \n_\alpha\n_\beta S+2\left(\n^\rho\n_{(\alpha} \xi\right)\n_{\beta)}\n_\rho S\nonumber\\
&&+\xi \n_\alpha\n_\beta \Box S+\zeta_4\left(\n_{(\alpha}R_{\beta)\rho}\right)\n^\rho S-\xi\left(\n^\lambda R_{\lambda(\alpha\beta)\rho}\right)\n^\rho S\nonumber\\
&&-2\left(\n^\lambda\xi\right)R_{\lambda(\alpha\beta)\rho}\n^\rho S
+2\left(\n_{(\alpha}\xi\right)\n_{\beta)}\Box S=8 \pi G T_{\alpha\beta}\,.\label{D4covequ}
\ees
As always, the Friedmann equation is then obtained from the $(00)$ component of this equation. 
This  covariant result is the necessary starting point when performing perturbation theory. Given the
large number of terms in \eq{D4covequ},
for obtaining the equations determining the background evolution, the fastest route is actually
to work directly in FRW.
Given the conformal properties of the $\Delta_4$ operator, to derive the cosmological equations it is convenient to use conformal time, so $ds^2=a^2(-d\eta^2+d\vx^2)$.
We also use the notation ${\cal H}=\pa_{\eta}a/a$. We will write explicitly the derivative with respect to $\eta$, reserving again the prime for $d/dx$, with $x=\ln a$.
\Eq{barDP} allows us to compute immediately the expression of $\DP$ on a FRW metric. Using conformal time,  $\gmn=a^2(\eta)\emn$. Then \eq{barDP} shows  that, on a scalar $S$,
$a^4\DP=\pa_{\eta}^4$\, i.e.
\be\label{DPinFRW}
\DP S =\frac{1}{a^4}\, \frac{\pa^4 S}{\pa\eta^4}\, .
\ee
To compute the equations of motion in FRW we however need to keep also the lapse function, considering a metric of the form 
\be\label{metricN}
\gmn=a^2(\eta) (-N^2(\eta),1,1,1)\,  .
\ee
The explicit computations shows that on this metric
\be\label{DPlapse}
\DP S =\frac{1}{a^4N^4}\, \[ \pa_{\eta}^4 S- 6  \pa_{\eta}n \, \pa_{\eta}^3S
-4 \pa_{\eta}^2n\, \pa_{\eta}^2S-\pa_{\eta}^3n\, \pa_{\eta}S\]
\, ,
\ee
where $\pa_{\eta}n\equiv (\pa_{\eta}N)/N$, i.e. $n=\log N$. In the above expression we have neglected all terms that contains products of more than one derivative of $n$, e.g $(\pa_{\eta}n)^2$, $\pa_{\eta}n \pa^2_{\eta}n$, etc. Indeed, to derive the Friedman equation we must take the variation of the action wrt to the lapse function $N$ and then set $N=1$, so $\pa_{\eta}n=0$. Then, all terms with  products of more than one derivative of $n$ give zero in the variation.
In the limit $N=1$ this expression  correctly reduces to
\eq{DPinFRW}.
This result can also be derived more elegantly defining a new conformal time $\tilde{\eta}$ from 
$Nd\eta=d\tilde{\eta}$. Then $ds^2=a^2(-d\tilde{\eta}^2+d\vx^2)$ is again conformal, and therefore
$a^4\DP =\pa^4/\pa\tilde{\eta}^4$. Since $\pa/\pa\tilde{\eta}=(d\eta/d\tilde{\eta})\pa/\pa\eta=(1/N)\pa/\pa\eta$, we get
\be
a^4\DP S=\(\frac{1}{N}\, \frac{\pa}{\pa\eta}\)^4 S\, .
\ee
Computing the derivatives and retaining only terms with at most one derivative of $N$ gives back \eq{DPlapse}. This also makes it clear why all terms  involving ${\cal H}$ `miracolously' cancel in a brute-force computation of \eq{DPlapse} from \eq{defDP}.\footnote{For instance, the term $\pe S$ in \eq{DPlapse}, beside being multiplied by  $\pa_{\eta}^3n$, in the intermediate steps of the computation is also multiplied by terms ${\cal H}^2$, ${\cal H}\pe {\cal H}$,
${\cal H}^2\pe n$ and $\pe{\cal H}\pe n$, and similarly for the terms $\pe^2 S$ and $\pe^3 S$.}

Restricting  to time-dependent fields in the metric (\ref{metricN}), the action (\ref{actDelta4}) becomes 
\be\label{actDelta4L}
S_{\Delta_4}=\frac{\mplr^2}{2}\int d\eta d^3x \, {\cal L}\, ,
\ee
where
\be\label{L1}
{\cal L}=\frac{6a^2}{N}\(\pa_{\eta}{\cal H}+{\cal H}^2-{\cal H}\frac{\pe N}{N}\) (1-\mu S+\xi)
-\frac{\xi}{N^3}\(  \pa_{\eta}^4 S- 6  \pa_{\eta}n \, \pa_{\eta}^3S
-4 \pa_{\eta}^2n\, \pa_{\eta}^2S-\pa_{\eta}^3n\, \pa_{\eta}S\)\, .
\ee
The variation with respect to $\xi$, at $N=1$, gives $\DP S=R$, i.e.
\be\label{pa4S}
\pe^4 S=6a^2 (\pe {\cal H}+{\cal H}^2)\, ,
\ee
while the variation with respect to $S$, again at $N=1$, gives\footnote{Notes that this is not true on a generic background. In FRW, with $N=1$, in the term proportional to $\xi$ in \eq{L1} only $\xi\pa_{\eta}^4 S$, survives, which can be trivially integrated by parts to give $S\pa_{\eta}^4 \xi$. However, in a general background $\xi\DP  S$ does not integrate by parts to $S\DP\xi$.}
\be
\xi=-\mu S\, .
\ee
The variation with respect to $N$, at $N=1$, can be obtained setting directly $\xi=-\mu S$ in \eq{L1}, i.e. using as a Lagrangian
\be
{\cal L}=\frac{6a^2}{N}\(\pa_{\eta}{\cal H}+{\cal H}^2-{\cal H}\frac{\pe N}{N}\) (1-2\mu S)
+\frac{\mu S}{N^3}\(  \pa_{\eta}^4 S- 6  \pa_{\eta}n \, \pa_{\eta}^3S
-4 \pa_{\eta}^2n\, \pa_{\eta}^2S-\pa_{\eta}^3n\, \pa_{\eta}S\)\, .
\ee
The variation wrt to $N$ gives (adding also the matter action)
\be
\frac{\d {\cal L}}{\d N}-\pe \(\frac{\d {\cal L}}{\d\, \pe N}\)+
\pe^2 \(\frac{\d {\cal L}}{\d\, \pe^2 N}\)
-\pe^3 \(\frac{\d {\cal L}}{\d\, \pe^3 N}\)=
\frac{2}{\mplr^2}\, \frac{\d {\cal L_{\rm matter}}}{\d N}\, .
\ee
where, after the variation, we set $N=1$. 
This 
gives the modified Friedman equation,
\be\label{Frimod1}
(1-2\mu S)a^2{\cal H}^2+\frac{\mu}{6}\[2\pe S\pe^3 S-(\pe^2 S)^2-12a^2{\cal H}\pe S\]
=\frac{8\pi G}{3}a^4\rho\, .
\ee
We have checked that this result agrees with that obtained directly from the $(00)$ component of the covariant equation of motion (\ref{D4covequ}).
We now introduce 
\be\label{defU}
U=\frac{1}{a^2}\pe^2S\, ,
\ee
and, in  \eq{Frimod1}, we express $\pe^2S$ and $\pe^3S$ in terms of $U$ and $\pe U$.
After  defining, as in Sect.~\ref{sect:nonmin},
$V(x)=H_0^2 S(x)$, $\gamma =m^2/(9H_0^2)$, $h(x)=H(x)/H_0$ and $\zeta(x)=h'(x)/h(x)$, \eq{Frimod1} reads
\be\label{fun1}
h^2(x)=\frac{\Omega(x) +(\gamma/4) U^2}{ 1+\gamma[ -3V'-3 V +(1/2)V'(U'+2U) ]}\, ,
\ee
In terms of the these dimensionless variables the equation $\pe^2S=a^2 U$, which follows from \eq{defU}, reads
\be\label{fun2}
V''+(1+\zeta)V'=h^{-2}U\, ,
\ee
while \eq{pa4S}, expressed in terms of $U$, reads
\be\label{fun3}
U''+(5+\zeta)U'+(6+2\zeta)U=6(2+\zeta)\, .
\ee
\Eqst{fun1}{fun3} are the fundamental equations for studying the background cosmology.
Again, it is also useful to rewrite the equations  trading $V$ for $W=h^2V$. Then \eq{fun1} becomes
\be\label{h2Y}
h^2(x)=\Omega(x)+\gamma Y\, ,
\ee
where 

\begin{figure}[t]
\centering
\includegraphics[width=0.48\columnwidth]{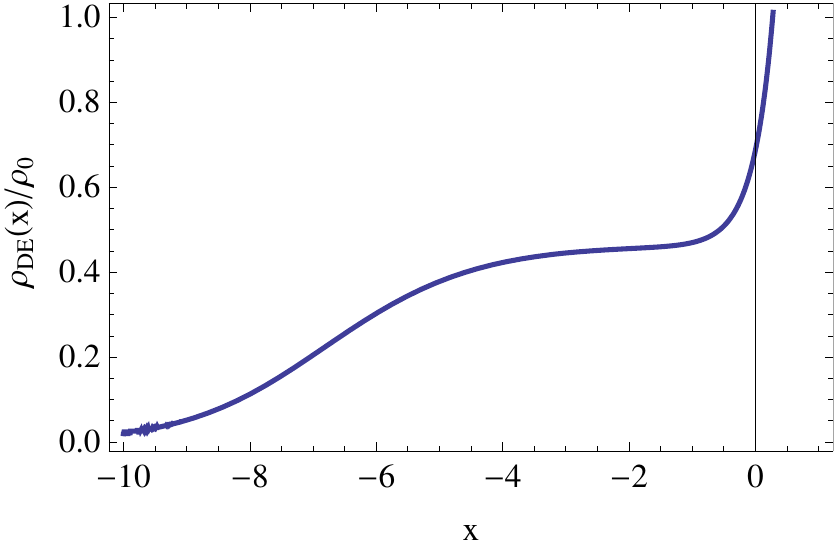}
\includegraphics[width=0.48\columnwidth]{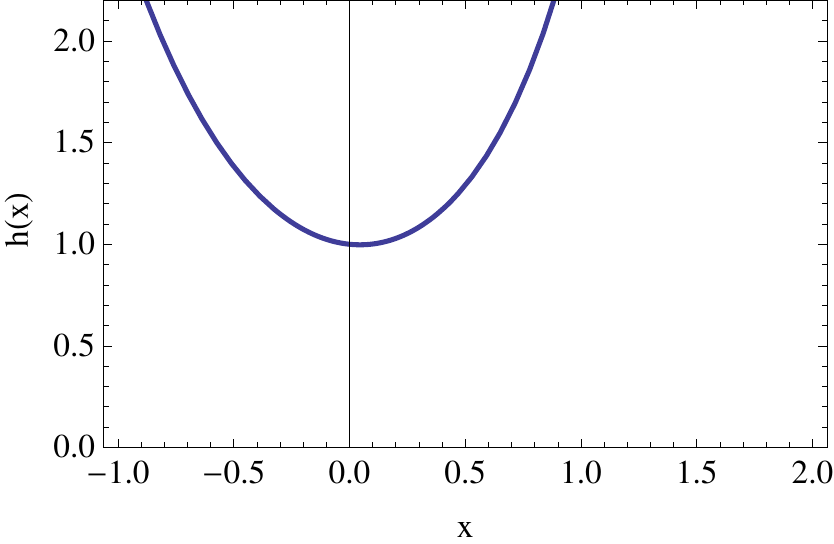}
\caption{\label{fig:rhoh} 
Left: the dark energy density $\rde(x)/\rho_0$.
Right: $h(x)$.
}
\end{figure}

\begin{figure}[t]
\centering
\includegraphics[width=0.48\columnwidth]{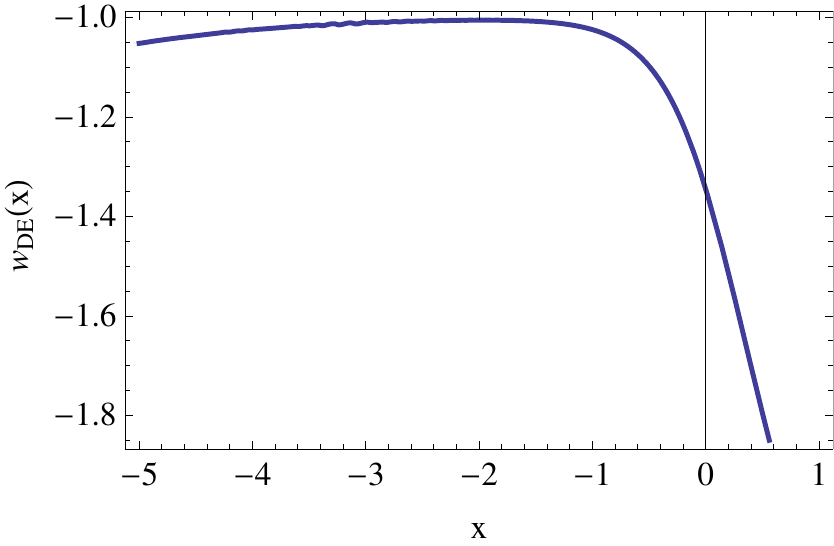}
\includegraphics[width=0.48\columnwidth]{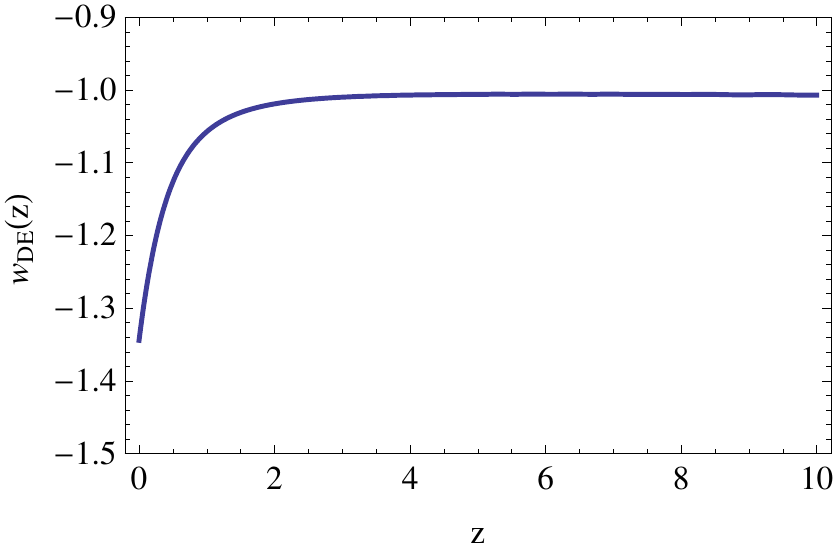}\\
\includegraphics[width=0.48\columnwidth]{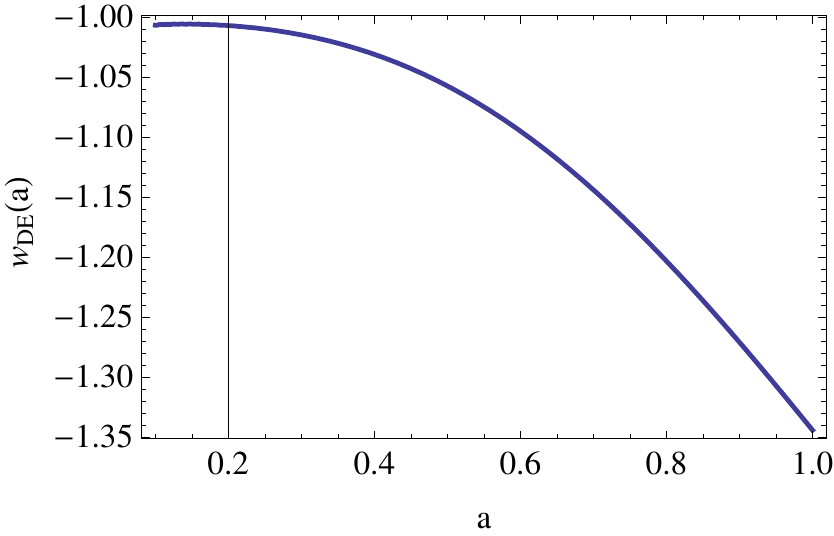}
\caption{\label{fig:w}   The EOS parameter of dark energy, $w_{\rm DE}$, as a function of $x=\ln a$ (top left), of redshift $z$ (top right) and  of scale factor $a$ (bottom). }
\end{figure}

\be\label{defY}
Y=\frac{1}{2}W'(6-U'-2U)+W(3-6\zeta+\zeta U'+2\zeta U)+\frac{1}{4}U^2\, ,
\ee
quite similar to eqs.~(31) and (33) of \cite{Maggiore:2014sia} (or (3.2) and (3.5) of \cite{Dirian:2014ara}), except that $U'$ is replaced by $U'+2U$. We see that
$\gamma Y$ plays the role of the effective dark energy fraction, $\ode (x)=\gamma Y$, as in \eq{defode}.
\Eq{fun2} is replaced by
\be\label{eqW}
W''+(1-3\zeta)W'+2(\zeta^2-\zeta-\zeta')W=U\, .
\ee
In this case the fundamental equations are \eqst{fun3}{eqW}. The solutions of the homogeneous equation
\be\label{fun3hom2}
U''+(5+\zeta_0)U'+(6+2\zeta_0)U=0\,
\ee
are $U=e^{\a_{+} x}$ and $U=e^{\a_{-} x}$ with $\a_{+}=-2$ and $\a_{-} =-(3+\zeta_0)$, which are both negative in all three eras, and indeed whenever 
$\zeta_0>-3$, which is always the case in the early Universe, where $\zeta_0\geq -2$ because of the contribution of radiation. These solutions induce inhomogeneous solutions of \eq{eqW} with the same behavior. Furthermore, there are solutions for $W$ corresponding to the the homogeneous solution of \eq{eqW} with $\zeta=\zeta_0$, 
\be\label{eqWhom}
W''+(1-3\zeta_0)W'+2(\zeta_0^2-\zeta_0)W=0\, .
\ee
These are given by $W=e^{\beta_{+} x}$ and  $W=e^{\beta_{-} x}$ with $\beta_{+}=2\zeta_0$ and $\beta_{-}=-1+\zeta_0$. Again, $\beta_{-}$ is negative in all three eras, while  $\beta_{+}$ is negative in RD and MD and vanishes, corresponding to a constant solution, in dS. Thus, there is no growing mode and the cosmological evolution is stable.

We now integrate \eqst{fun1}{fun3} numerically. As before, we need to compute explicitly $\zeta$ in terms of $U,V,U'$ and $V'$.  This gives
\be
\zeta= \frac{1}{2(1-3 \gamma V)}
\left\{ h^{-2}\Omega'+\g\[h^{-2}U(3-U)+V'(2U'+4U-6)\]\right\}\, .
\ee
Then \eqs{fun2}{fun3} become a closed system for $U,V$ that can be integrated, and the solution can then be simply plug into \eq{fun1} to get $h(x)$.  Fixing $\oma=0.313245$ we get $\gamma =0.318267$.
The result of the numerical integration for the functions $\rde(x)$ and $h(x)$ is shown in Fig.~\ref{fig:rhoh}, while
$w_{\rm DE}$ is shown in Fig.~\ref{fig:w}.

The value of $w_{\rm DE}(x)$ today is $w_{\rm DE}(0)\simeq -1.34$.  For a long range of red-shifts $w_{\rm DE}$ is very close  to $-1$ (e.g., at $z=6$, $w_{\rm DE}\simeq -1.006$), i.e. $\rde$ is a approximately constant during MD, as we also see from Fig.~\ref{fig:rhoh}.
Then $w_{\rm DE}$  drops toward more phantom values when the DE starts to dominate, leading to a fast growth of $\rde(x)$ and $h(x)$.

To assess whether this prediction for $w_{\rm DE}(z)$ is consistent with cosmological observations, a full analysis of the cosmological perturbations, as well as the corresponding parameter estimation in the $R\DP^{-1}R$ model, is in principle necessary.  However,  the right-bottom panel of 
Fig.~\ref{fig:w} shows that,  at least at the level of background evolution, near the present epoch a linear fit of the form (\ref{w0wa}) is appropriate.
Fitting our results to \eq{w0wa} we get
\be\label{fitw0wa}
w_0\simeq -1.31\, ,\qquad w_a=0.49\, .
\ee 
Comparing the values in \eq{fitw0wa} with Fig.~4  of the 2015 Planck dark energy paper \cite{Ade:2015rim},
or comparing directly our   Fig.~\ref{fig:w}  (upper-right panel) for $w(z)$ with  Fig.~5 of  \cite{Ade:2015rim} we see that, if one combines Planck CMB data with BAO, SNe and $H_0$ measurements, the predictions of the $\Delta_4$ model are excluded at more than 95\% c.l. (and possibly, extrapolating from the contours of the figures in \cite{Ade:2015rim}, at about 99\% c.l.). Thus, unless the inclusion of cosmological perturbations changes substantially the picture, the  $\Delta_4$ model  is basically excluded by the data.

\section{Conclusions and further directions}\label{sect:Concl}

In this paper we have continued our exploration of the landscape of possible viable nonlocal IR modifications of GR. From our previous works, we have been lead to focus on terms in the action in which a nonlocal operator is sandwiched between two Ricci scalars. The simplest option, the RR model (\ref{RR}) where $R\Box^{-2}R$ appears in the action, by itself fits well the data. However,  once we use the most recent 2015 Planck data, it is significantly disfavored compared to $\Lambda$CDM and  to the RT 
nonlocal model (\ref{RT})\cite{Diraninprep}. In contrast, the RT model and  $\Lambda$CDM fit the data equally well. Since the RT model is 
in a sense a nonlinear extension of the RR model, we were lead to examine other possible forms of such nonlinear extensions. Of course in principle there is an infinity of choices, as often happens in model building. Symmetries are however a powerful guiding principle in model building, and we have then chosen to explore two possibilities that might be indications of an underlying conformal symmetry. We have found that the first model, defined by the action (\ref{xiRR}), works very well, while a model constructed with the Paneitz operator, \eq{actDelta4}, seems ruled out.

The model (\ref{xiRR}) seems indeed to enjoy several positive features, even compared to the other nonlocal model that works well, i.e. the RT model. Indeed, it is defined in terms of a relatively simple action. As we have discussed, its cosmological evolution is stable both in an early de~Sitter inflationary era, as well as in the subsequent RD and MD epochs, while the RT model is only stable in RD and MD, and must therefore be eventually embedded in some high-energy modification at the inflationary scale. The specific form of the nonlocal term is protected by conformal symmetry, which gives a motivation for discarding the possibility of adding also a mass term, $-\Box\ra -\Box +m^2$, which is otherwise in principle possible for the RR and RT models. At the level of comparison with the data, a more detailed study of its cosmological perturbations is necessary to make quantitative assessment, and a Bayesian comparison with $\Lambda$CDM. However, from the background evolution,  we see that the deviation of this model from $\Lambda$CDM are very tiny, at the level of about $1\%$. We expect that, once the full apparatus of computing the cosmological perturbations and implementing them in a modified Boltzmann code will be developed,   its predictions will be intermediate between those of $\Lambda$CDM and those of the RT model, and possibly near the limit of resolution of future missions such as Euclid. 

It is also interesting to observe that nonlocal models featuring $R^2$ terms in the action, which can explain DE in the recent epoch, can be naturally connected with Starobinski inflation at high energies. For instance, one could generalize \eq{6RR} into 
\be\label{6RRStar}
S_{\rm conf}=\frac{\mplr^2}{2}\int d^4x\, \sqrt{-g}\, 
\left\{ R+\frac{1}{6M_{\rm S}^2} R\[ 1- \frac{\Lambda_{\rm S}^4}{(-\Box+  \frac{1}{6}R)(-\Box +\frac{1}{6} R)}\] R\right\}\, ,
\ee
where $M_{\rm S}\simeq 10^{13}$~GeV is the mass scale of the Starobinski model and $\Lambda_{\rm S}^4=6M_{\rm S}^2\mu=M_{\rm S}^2m^2$,
so $\Lambda_{\rm S}= (M_{\rm S}m)^{1/2}$.
We found in \eq{m085} that, setting $\oma=0.313$, we get $m\simeq 0.85 H_0$ (of course,  the precise numerical coefficient depends on the value of $\oma$, that should eventually be obtained by performing parameter estimation of the model). 
Then, numerically, $\Lambda_{\rm S}= {\cal O}(10^{-6})\, {\rm eV}$. Similar interpolations could be done for the RR model and (directly at the level of equations of motions) for the RT model.
 
At sufficiently high energies   or curvatures we have
$R (\Lambda_{\rm S}^4/\Box^2) R\ll R^2$.\footnote{Of course, the actual numerical estimate must take into account that the $\iBox$ operator actually depends on the whole past history of the system.} In this regime the nonlocal term is much smaller than the local one, and we recover the Starobinski model. This means that, at $E\sim M$, we have the standard Starobinski inflationary phase. After reheating, when the energy and curvature drop below the scale $M$, the local $R^2$ term becomes negligible with respect to the Einstein-Hilbert term. Finally, in the recent epoch, say $z<10$, the nonlocal term gradually starts to become important, and the evolution is the one obtained from the nonlocal model (\ref{xiRR}). Thus, an effective action such as 
(\ref{6RRStar}) nicely interpolates between  inflation in the primordial Universe and accelerated expansion in the recent epoch. It would be interesting to understand if such a model could reflect a renormalization-group flow in a gravity theory with $R^2$ term, as tentatively discussed in \cite{Maggiore:2015rma}.

\vspace{5mm}
\noindent
{\bf Acknowledgments.} We thank Ermis Mitsou for comments on the manuscript, and Eugenio Bianchi for early discussions on the relation to the Starobinski model.
The work of G.Cusin, S.Foffa and M.Maggiore is supported by the Fonds National Suisse. The work of M.Maggiore is supported by the SwissMap NCCR.

\appendix
\section{Causality in  nonlocal theories}\label{sect:causality}

The discussion of the causality in nonlocal theories involves some subtle point. Even if these issues have already  been correctly discussed several times in the literature,  see e.g. \cite{Tsamis:1997rk,Deser:2007jk,Barvinsky:2011rk,Deser:2013uya,Ferreira:2013tqn,Foffa:2013sma}, we find useful to summarize again here the issue, in order to clarify some confusion which occasionally resurfaces.

The crucial, and possibly confusing, point is that  an effective nonlocal theory must be treated differently from  a fundamental  field theory. For instance, if we consider  a nonlocal action and we take naively its variation in the standard way, we unavoidably obtain acausal equations of motion. The simplest example, discussed in \cite{Foffa:2013sma},
is given by a nonlocal term in the action of a scalar field, of the form
$\int dx \phi\iBox\phi$, where $\phi$ is some scalar field, and $\iBox$ is defined with respect to some Green's function $G(x;x')$. Taking the variation with respect to $\phi(x)$ we get
\bees
&&\frac{\d}{\d\phi(x)}\int dx' \phi(x') (\iBox\phi )(x')=
\frac{\d}{\d\phi(x)} \int dx' dx'' \phi(x') G(x';x'') \phi(x'')\nn\\
&&=\int dx' [G(x;x')+G(x';x)] \phi(x')\, . \label{symGreen}
\ees
We see that the variation of the action automatically symmetrizes the Green's
function. It is therefore impossible to obtain in this way a retarded Green's function in the equations of motion, since $G_{\rm ret}(x;x')$ is not symmetric under $x\leftrightarrow x'$;  rather 
$G_{\rm ret}(x';x)=G_{\rm adv}(x;x')$. So, even if we start from a retarded Green's function in the action, we get a combination of retarded and advanced Green's function in the equation of motion. The same unavoidably happens if we formally take the variation of a nonlocal gravity action such as that of  the RR model  \cite{Maggiore:2013mea,Foffa:2013sma,Zhang:2016ykx}. This is indeed one of the reason why the action of a {\em fundamental} field theory must be local.

However quantum effective actions can, and in fact almost unavoidably are, nonlocal. Indeed, a nonlocal quantum effective action just describes, in coordinate space, the running of coupling constants that is more commonly described in momentum space. For instance, at  one-loop level the running of the electric charge in QED can be described in coordinate space by the one-loop quantum effective action~\cite{Barvinsky:1987uw} (see also \cite{Dalvit:1994gf,Lombardo:1996gp})
\be\label{qed}
S_{\rm eff}=-\frac{1}{4}\int d^4x\, F_{\mu\nu}\frac{1}{e^2(\Box)}F^{\mu\nu}\, ,
\ee
where
\be
\frac{1}{e^2(\Box)}=\frac{1}{e^2(\mu)}-\beta_0\log\(\frac{-\Box}{\mu^2}\)\, .
\ee
Here $\mu$ is the renormalization scale, $e(\mu)$ is the renormalized charge at the scale $\mu$  and, for a single massless fermion, $\beta_0=1/(12\pi^2)$. The logarithm of the d'Alembertian can be defined for instance from 
\be
\log\(\frac{-\Box}{\mu^2}\)=\int_0^{\infty}dm^2\, \[\frac{1}{m^2+\mu^2}-
\frac{1}{m^2-\Box}\]\, .
\ee
Another particularly famous example of nonlocal quantum effective action is the  Polyakov action in $D=2$ space-time dimensions,
\be\label{SPolyakov}
S_{\rm P}=-\frac{N}{96\pi}\int d^2x\sqrt{-g}\, R\frac{1}{\Box}R\, .
\ee
The Polyakov action is the quantum effective action obtained by integrating out the massless matter fields in 2-dimensional gravity coupled to matter, and can also be obtained integrating the conformal (or trace) anomaly. In $D=2$
the conformal anomaly takes the form (see e.g. \cite{Birrell:1982ix}) 
\be\label{traceT}
\langle T^{\mu}_{\mu}\rangle=\frac{N}{24\pi} R\,,
\ee
where $N=N_S+N_F$ is the total number of massless scalar and Dirac fermion fields.\footnote{The coefficient $N$ in front of  $S_{\rm anom}$  becomes $N-25$ if one also takes into account the metric fluctuations themselves, beside the fluctuations due to matter fields.} The energy-momentum tensor obtained taking the variation of $S_{\rm P}$ has indeed a trace equal to the right-hand side of \eq{traceT}. Thus, 
\eq{SPolyakov}
can  be added to the classical Einstein-Hilbert action (which in $D=2$ is just a topological invariant), to provide an effective action which takes into account  the quantum effect due to loops of massless particles.

Similarly, in $D=4$, starting from gravity coupled to massless matter fields and integrating out the massless matter fields, one obtains the  nonlocal action (see e.g. \cite{Birrell:1982ix,Buchbinder:1992rb,Shapiro:2008sf} for pedagogical introductions)
\be\label{SanomD4}
S_{\rm anom}=-\frac{1}{8}\int d^4x\sqrt{-g}\(E-\frac{2}{3}\Box R\) \Delta_4^{-1}
\[ b' \(E-\frac{2}{3}\Box R\)-2b C^2\]\, ,
\ee
where $E$ is the Gauss-Bonnet term,  $C^2=C_{\mu\nu\rho\sigma}C^{\mu\nu\rho\sigma}$ is the square of the Weyl tensor, $\Delta_4$ is the Paneitz operator (\ref{defDP}), and the coefficients $b,b'$ depends on the number of scalar, vector and tensor massless fields  integrated out. Again, the corresponding energy-momentum tensor reproduces the conformal anomaly, and the action (\ref{SanomD4}) can equivalently be obtained integrating the anomaly. Thus, \eq{SanomD4} is  called the anomaly-induced effective action.

Of course, if one would naively take the variation of these nonlocal actions for deriving the corresponding Euler-Lagrange equations, one would find again acausal equations of motions, with symmetrized Green's function.
However, it is obvious that this cannot be a sign of a genuine physical acausality, since these nonlocal effective actions are just a way to express, with an action that can be used at tree level, the result of a  one-loop quantum computation in fundamental theories, like QED or gravity with massless matter fields, which are local and  causal. 
The resolution of this apparent paradox is that
the equations of motion derived from the quantum effective action are no longer equation for the classical fields involved, say a classical field $\phi$ or the classical metric $\gmn$. Rather, they are the  equations of motion obeyed by the vacuum expectation values of the corresponding operators, $\cav\hat{\phi}\vac$ or $\cav\hat{g}_{\mu\nu}\vac$. However, now we must specify whether we consider the in-in or the in-out expectation values, i.e. 
$\langle 0_{\rm in}|\hat{\phi} |0_{\rm in}\rangle$ or $\langle 0_{\rm out}|\hat{\phi }|0_{\rm in}\rangle$.

The classical equation for the in-out expectation values  correspond to a diagrammatic expansion of  the usual Feynman path  integral. Then one finds that the in-out expectation values indeed obey nonlocal and acausal equations of motion, where the nonlocal operators, such as $\iBox$, are defined with the Feynman Green's function. Of course, there is nothing wrong with it. The in-out matrix element are not observable quantities, but just auxiliary  objects which enter in intermediate steps in the computation of scattering amplitudes. Furthermore, even if an operator such as $\hat{\phi}$ or  $\hat{g}_{\mu\nu}$ is  hermitean, its in-out matrix element are complex. In particular, this makes it impossible to interpret 
$\langle 0_{\rm out}|\hat{g}_{\mu\nu}|0_{\rm in}\rangle$ as an effective metric.  The in-out matrix elements do not have to obey causal equations (and indeed the Feynman propagator, which is  acausal, enters everywhere in QFT computations). 
In contrast, the in-in matrix elements are the semiclassical quantities that are in principle observables (and are real, if the corresponding operators are hermitean). The equations of motion for the in-in expectation values correspond to
a diagrammatic expansion of the Schwinger-Keldysh path integral, which automatically provides nonlocal but causal 
equations  \cite{Jordan:1986ug,Calzetta:1986ey}, involving only the  retarded propagator.

Thus, nonlocal actions such as (\ref{RR}) or (\ref{6RR}), interpreted as quantum effective actions, correspond to causal theories, exactly as, for instance, the one-loop effective action for QED (\ref{qed}) or the Polyakov action (\ref{SPolyakov}).

Non-local but causal equations can  also emerge from a purely classical averaging procedure, when one separates the dynamics of a system into a long-wavelength and a short-wavelength part. Here the mechanism ensuring causality is even simpler. Suppose that a system has a degree of freedom $\phi$ that we wish to eliminate, coupled to a set of degrees of freedoms, that we denote collectively as $\Psi$, that we wish to retain. Classically, $\phi$ could satisfy an equation, say, of the form $\Box\phi=j(\Psi)$. This equation is then solved as $\phi=\Box^{-1}_{\rm ret} j(\Psi)$, where the retarded propagator is selected by causality, as always in such classical computations. This solutions is then re-injected in the equations for the remaining degrees of freedom $\Psi$, which were coupled to $\phi$, e.g. $\Box\Psi=f(\phi)$. As a consequence, $\Psi$ now satisfies nonlocal but causal equations. An example of this form, in the context of cosmological perturbation theory, is discussed in  \cite{Carroll:2013oxa}. If this is the mechanism behind nonlocal equations such as (\ref{RT}), then again the causality of the equations is automatically assured. In this case actions such as (\ref{RR}) or (\ref{xiRR}) can be interpreted simply as convenient tools for generating nonlocal equations of motions that are automatically covariant. However, in this case the fundamental quantity would be the equation of motion, rather than the action, in which $\iBox$ is then taken to be the retarded propagator.

\bibliographystyle{utphys}
\bibliography{myrefs_massive}

\end{document}